\documentclass[%
twocolumn,
superscriptaddress,
altaffilletter,
showpacs,
nobibnotes,
 amsmath,amssymb,
 aps,prc,
floatfix,
]{revtex4-1}

\usepackage[utf8]{inputenc}
\usepackage{graphicx}
\usepackage{dcolumn}
\usepackage{bm}
\usepackage{hyperref}
\usepackage{hepparticles} 
\usepackage{hepnames} 
\usepackage[alsoload=hep]{siunitx}
\usepackage{natbib}
\usepackage{color}

\begin{document}


\title{Determination of the Spin Triplet $p\Lambda$ Scattering Length from the Final State Interaction in the $\vec{p}p \rightarrow pK^{+}\Lambda$ Reaction}

\author{F. Hauenstein}
\email{f.hauenstein@fz-juelich.de}
\affiliation{Institut f\"{u}r Kernphysik, Forschungszentrum J\"{u}lich, 52428 J\"{u}lich, Germany}
\affiliation{Friedrich-Alexander-Universit\"{a}t Erlangen-N\"{u}rnberg, 91058 Erlangen, Germany}

\author{E. Borodina}
\affiliation{Institut f\"{u}r Kernphysik, Forschungszentrum J\"{u}lich, 52428 J\"{u}lich, Germany}
\author{H. Clement}
\affiliation{Physikalisches Institut der Universität T\"{u}bingen, Auf der Morgenstelle 14, 72076 T\"{u}bingen, Germany}
\affiliation{Kepler Center for Astro and Particle Physics, University of T\"{u}bingen, Auf der Morgenstelle 14, 72076 T\"{u}bingen, Germany}
\author{E. Doroshkevich}
\altaffiliation{current address: Institute for Nuclear Research, Moscow 117312, Russia}
\affiliation{Physikalisches Institut der Universität T\"{u}bingen, Auf der Morgenstelle 14, 72076 T\"{u}bingen, Germany}
\affiliation{Kepler Center for Astro and Particle Physics, University of T\"{u}bingen, Auf der Morgenstelle 14, 72076 T\"{u}bingen, Germany}
\author{R. Dzhygadlo}
\altaffiliation{current address: Hadron Physics 1, Gesellschaft f\"{u}r Schwer\-io\-nen\-for\-schung (GSI), Darmstadt, Germany}
\affiliation{Institut f\"{u}r Kernphysik, Forschungszentrum J\"{u}lich, 52428 J\"{u}lich, Germany}
\author{K. Ehrhardt}%
\affiliation{Physikalisches Institut der Universität T\"{u}bingen, Auf der Morgenstelle 14, 72076 T\"{u}bingen, Germany}
\affiliation{Kepler Center for Astro and Particle Physics, University of T\"{u}bingen, Auf der Morgenstelle 14, 72076 T\"{u}bingen, Germany}
\author{W. Eyrich}
\affiliation{Friedrich-Alexander-Universit\"{a}t Erlangen-N\"{u}rnberg, 91058 Erlangen, Germany}
\author{W. Gast}
\affiliation{Institut f\"{u}r Kernphysik, Forschungszentrum J\"{u}lich, 52428 J\"{u}lich, Germany}
\author{A. Gillitzer}
\affiliation{Institut f\"{u}r Kernphysik, Forschungszentrum J\"{u}lich, 52428 J\"{u}lich, Germany}
\author{D. Grzonka}
\affiliation{Institut f\"{u}r Kernphysik, Forschungszentrum J\"{u}lich, 52428 J\"{u}lich, Germany}
\author{J. Haidenbauer}
\affiliation{Institut f\"{u}r Kernphysik, Forschungszentrum J\"{u}lich, 52428 J\"{u}lich, Germany}
\affiliation{Institut for Advanced Simulation and J\"{u}lich Center for Hadron Physics, Forschungszentrum J\"{u}lich, 52428 J\"{u}lich, Germany}
\author{C. Hanhart}
\affiliation{Institut f\"{u}r Kernphysik, Forschungszentrum J\"{u}lich, 52428 J\"{u}lich, Germany}
\affiliation{Institut for Advanced Simulation and J\"{u}lich Center for Hadron Physics, Forschungszentrum J\"{u}lich, 52428 J\"{u}lich, Germany}
\author{S. Jowzaee}
\affiliation{Institut f\"{u}r Kernphysik, Forschungszentrum J\"{u}lich, 52428 J\"{u}lich, Germany}
\affiliation{Institute of Physics, Jagellonian University, PL-30-348 Cracow, Poland}
\author{K. Kilian}
\affiliation{Institut f\"{u}r Kernphysik, Forschungszentrum J\"{u}lich, 52428 J\"{u}lich, Germany}
\author{P. Klaja}
\affiliation{Institut f\"{u}r Kernphysik, Forschungszentrum J\"{u}lich, 52428 J\"{u}lich, Germany}
\affiliation{Friedrich-Alexander-Universit\"{a}t Erlangen-N\"{u}rnberg, 91058 Erlangen, Germany}
\author{L. Kober}
\affiliation{Friedrich-Alexander-Universit\"{a}t Erlangen-N\"{u}rnberg, 91058 Erlangen, Germany}
\author{M. Krapp}
\affiliation{Friedrich-Alexander-Universit\"{a}t Erlangen-N\"{u}rnberg, 91058 Erlangen, Germany}
\author{M. Mertens}
\altaffiliation{current address: University Duisburg-Essen, 45141 Essen, Germany}
\affiliation{Institut f\"{u}r Kernphysik, Forschungszentrum J\"{u}lich, 52428 J\"{u}lich, Germany}
\author{P. Moskal}
\affiliation{Institute of Physics, Jagellonian University, PL-30-348 Cracow, Poland}
\author{J. Ritman}
\affiliation{Institut f\"{u}r Kernphysik, Forschungszentrum J\"{u}lich, 52428 J\"{u}lich, Germany}
\affiliation{J\"{u}lich Aachen Research Alliance, Forces and Matter Experiments (JARA-FAME), Germany}
\affiliation{Experimentalphysik I, Ruhr-Universit\"{a}t Bochum, 44780 Bochum, Germany}
\author{E. Roderburg}
\affiliation{Institut f\"{u}r Kernphysik, Forschungszentrum J\"{u}lich, 52428 J\"{u}lich, Germany}
\author{M. R\"{o}der}
\affiliation{Institut f\"{u}r Kernphysik, Forschungszentrum J\"{u}lich, 52428 J\"{u}lich, Germany}
\affiliation{Corporate Development, Forschungszentrum J\"{u}lich, 52428 J\"{u}lich, Germany}
\author{W. Schroeder}
\affiliation{Corporate Development, Forschungszentrum J\"{u}lich, 52428 J\"{u}lich, Germany}
\author{T. Sefzick}
\affiliation{Institut f\"{u}r Kernphysik, Forschungszentrum J\"{u}lich, 52428 J\"{u}lich, Germany}
\author{P. Wintz}
\affiliation{Institut f\"{u}r Kernphysik, Forschungszentrum J\"{u}lich, 52428 J\"{u}lich, Germany}
\author{P. W\"{u}stner}
\affiliation{Zentralinstitut f\"{u}r Engineering, Elektronik and Analytik, Forschungszentrum J\"{u}lich, 52428 J\"{u}lich, Germany}

\collaboration{COSY-TOF Collaboration}
\noaffiliation

\date{\today}

\begin{abstract}
The $\vec{p}p \rightarrow pK^{+}\Lambda$ reaction has been measured with the COSY-TOF detector at a beam momentum of $\SI{2.7}{\giga\eVperc}$. 
The polarized proton beam enables the measurement of the beam analyzing power by the asymmetry of the produced kaon ($A_N^{K}$). This observable allows the
$p\Lambda$ spin triplet scattering length to be extracted for the first time model independently from the final-state interaction in the reaction. The obtained value is $a_{t} = \SI[parse-numbers=false]{(-2.55 ^{+0.72}_{-1.39} {}_{stat.} \pm 0.6_{syst.} \pm 0.3_{theo.})}{\femto\meter}$. This value is compatible with theoretical predictions and results from model-dependent analyses.
\end{abstract}

\maketitle

The elementary hyperon-nucleon interaction is an essential ingredient
for microscopic few-body or many-body calculations. This concerns investigations 
of light hypernuclei \cite{Wirth2014},
like the hypertriton and 
${}^4_\Lambda {\rm He}$ \cite{Nogga2014}, 
as well as studies of neutron stars where hyperons are expected to be present in 
the core \cite{Lonardoni2015}. In particular, the interaction of the hyperons with the surrounding neutron matter has a crucial influence 
on the radius and mass of those stars.

Naturally, the reliability of pertinent calculations not only depends on the employed
few-body or many-body approaches, but also crucially on the interaction that 
is used as input. 
Therefore, it is unfortunate that so far only the bulk properties of the
hyperon-nucleon interaction are known from experiment. Specifically, with regard 
to the $\Lambda N$ interaction, there are no data from elastic scattering that would allow to pin down 
its spin dependence \footnote{The result for the reaction $K^{-}d \rightarrow \pi^{-}p\Lambda$ in \cite{Tan1969} is only sensitive to
spin triplet states if the interaction of the hyperon with one of the nucleons of the deuteron is dominated by s-wave.}. In addition, the chance is low for
any pertinent low-energy scattering data in the future due to the short hyperon lifetime which makes 
it difficult to prepare a hyperon beam or target.

An alternative is to study the
hyperon-nucleon interaction by the final-state interaction
(FSI) in high momentum transfer reactions such as $pp \rightarrow pK^{+}\Lambda$ as pointed out
in Ref. \cite{Gasparyan2004}. Indeed, the method developed and described in this reference allows
the extraction of the S-wave $p\Lambda$ scattering lengths from FSI effects with a
definite theoretical uncertainty and without model assumptions. 
The scattering length is determined directly from the shape of the 
$p\Lambda$ invariant mass spectrum. The spin-dependence of the FSI and
of the $p\Lambda$ interaction can be disentangled by considering specific
polarization states of the beam and/or target particles. 
For the spin-triplet component of the $p\Lambda$ interaction this can be
achieved by measuring the symmetric contribution of the beam 
analyzing power determined from the asymmetry of the produced kaon ($A^K_N$), which is accessible in an experiment with a polarized beam. 
For the first time the results of a model-independent determination of the $p\Lambda$ spin triplet scattering length 
are reported in the present paper. 

The $\vec{p}p \rightarrow pK^{+}\Lambda$ reaction has been exclusively measured with the COSY-TOF detector using a polarized proton beam with
a momentum of $\SI{2.7}{\giga\eVperc}$. The details of the detector system including the Straw-Tube-Tracker (STT) can be found in Refs.
\cite{Roeder2013, Jowzaee2016, Wintz2004, Roeder2011}. 

COSY-TOF covers the full phase space of the reaction. From Monte Carlo (MC) simulations the reconstruction efficiency times acceptance is determined to be
about \SI{15}{\percent} and almost constant over the full kinematic range. The average $p\Lambda$ invariant mass resolution is $\sigma_{m} \approx \SI{1.3}{\mega\eV}$. At the kinematic boarders
the resolution improves to $\sigma_{m} \approx \SI{0.5}{\mega\eV}$ due to the kinematic fitting. Possible effects on
the extraction procedure have been investigated by simulations and a dedicated analysis with a mass dependent resolution. The results show negligible differences  \cite{Hauenstein2014}.

The $pK^{+}\Lambda \rightarrow pK^{+}p\pi^{-} $ events have a clear signature of four tracks in the final state where two tracks stem from the target
and two tracks from the delayed weak decay of the $\Lambda$ hyperon. After applying a geometric fit, a kinematic fit is performed on each event with two
overconstraints.  

The final event selection criteria consists of constraints on (i) the reduced chi square of the kinematic fit, $\chi^{2}_{\text{kin}} / \text{NDF} < 5$, (ii) the minimum
distance $d_{\Lambda}$ between the production and the decay of the $\Lambda$, $d_{\Lambda} > \SI{3}{\centi\metre}$, and (iii) the minimum angle between
the directions of the decay proton and the $\Lambda$, $\angle(\Lambda, p) > 2^{\circ}$. Finally, 232,873 events are used for the further analysis.

Incorrectly reconstructed events with multiple primary tracks i.e. from multi-pion production are effectively removed by the applied criteria (i) $-$ (iii).
The remaining physical background stems from the $\vec{p}p \rightarrow pK^{+}\Sigma^{0} \rightarrow pK^{+}\Lambda\gamma$
reaction, which has nearly the same topology. However, MC simulations show that at this beam momentum the contribution from the $\Sigma^{0}$ production to the final event sample is below $\SI{1}{\percent}$
\cite{Hauenstein2014}. Therefore, it is neglected in the following analysis.

The methods for the determination of the beam polarization and the analyzing power $A_{N}^{K}$ are described in Refs.~\cite{Roeder2013,Hauenstein2014}.
The beam polarization for the measurement at $\SI{2.7}{\giga\eVperc}$ is $(77.9 \pm 1.2)\,\%$. It is determined from the asymmetry of elastic events
and the $pp$ analyzing power from the SAID partial 
wave analysis SP07 \cite{Arndt2007}. Systematic effects from different magnitudes of ``up'' and ``down'' beam polarization can be neglected in the following analysis as discussed in Ref. \cite{Roeder2013}.

The observable $A_{N}^{K}$ gives access to spin triplet $p\Lambda$ states due to its particular dependence on interference terms of kaon partial waves. Expanding $A_{N}^{K}$ in terms of associated Legendre polynomials $P^{m}_{l}$ gives \cite{Roeder2013,Hanhart2004}
\begin{eqnarray}
\frac{A_{N}^{K}(x,m_{p\Lambda})}{\Phi(s,m_{p\Lambda})}\frac{\mathrm{d}^2\sigma}{\mathrm{d}\Omega_{K}^{\mathrm{*}}\mathrm{d}m_{p\Lambda}} = \sum_{i=1}^{N} b_{i}(m_{p\Lambda}) P_{i}^{1}(x),
 \label{eq:generalexpression}
\end{eqnarray}
where $\Phi(s,m_{p\Lambda})$  is a phase space factor and $x = \cos\theta^{*}_{K}$. 
The coefficients $b_{1},b_{3},b_{5},..$ result from an interference of odd and even kaon partial waves
and in this case only the spin triplet $p\Lambda$ final states can contribute (for details see \cite{Gasparyan2004,Roeder2013,Hanhart2004}).
It turns out that only the two leading terms of Eq.~(\ref{eq:generalexpression}) are needed in the present analysis, hence $b_{1}(m_{p\Lambda})$ can be used for the determination of the spin triplet scattering length.

Using the parametrization 
\begin{equation}
\left| b_{1}(m_{p\Lambda}) \right| = \exp {\left[C_0 + \frac{C_1}{m_{p\Lambda}^{2} - C_2}\right]},
 \label{eq:fsiamplitudeparametrization}
\end{equation}
the spin triplet scattering length $a_{t}$ can be obtained by 
\begin{eqnarray}
a_{t}(C_1,C_2) &=& -\frac{\hbar c}{2} C_1 \, \sqrt{\frac{m_0^2}{m_{p} m_{\Lambda}}}\nonumber\\*
&\times& \sqrt{ \frac{(m_{\textrm{max}}^2 - m_0^2)}{(m_{\textrm{max}}^2-C_2) (m_0^2-C_2)^3}}
\label{eq:scatteringlengthformula}
\end{eqnarray}
where $m_0 = m_{\Lambda} + m_{p}$ and $m_{\textrm{max}} = m_0 + \SI{40}{\mega\eV}$. The latter value indicates the applied fit range as well as 
the upper limit of the dispersion integral from theory \cite{Gasparyan2004} to fulfill the requirement that the $p\Lambda$ system is dominantly in S-wave.
The dependence of the scattering length extraction on $m_{\textrm{max}}$ is included in the estimated theoretical uncertainty of $\SI{0.3}{\femto\meter}$ \cite{Gasparyan2004}.

The independence of Eq.~(\ref{eq:scatteringlengthformula}) from $C_0$ reflects the fact that only the shape of the FSI-enhancement is 
important to determine the scattering length \cite{Gasparyan2004}.
Therefore, the proportionality of $|b_{1}(m_{p\Lambda})|$ to the spin triplet scattering amplitude is sufficient to
determine $a_{t}$.

Since the kaon angular distribution is uniform \cite{Jowzaee2016}, $A_{N}^{K}$ is directly evaluated in terms of associated Legendre polynomials and bins of invariant mass $m_{p\Lambda}$, which gives
\begin{equation}
A_{N}^{K}(m_{p\Lambda}) = \alpha(m_{p\Lambda}) P_{1}^{1} + \beta(m_{p\Lambda}) P_{2}^{1}.
\end{equation} 
Combining this expansion with Eq.~(\ref{eq:generalexpression}) results in
\begin{equation} 
b_{1}(m_{p\Lambda}) = \alpha(m_{p\Lambda}) \cdot | \tilde{A}(m_{p\Lambda}) |^2 = \frac{\alpha(m_{p\Lambda})}{\Phi(s,m_{p\Lambda})} \frac{\mathrm{d}\sigma } { \mathrm{d}m_{p\Lambda} } .
\end{equation}
The two equations correspond to Eqs.~(5-7) in \cite{Roeder2013} with a different notation of the variables.

\begin{figure}[b]
\includegraphics[width=0.5\textwidth]{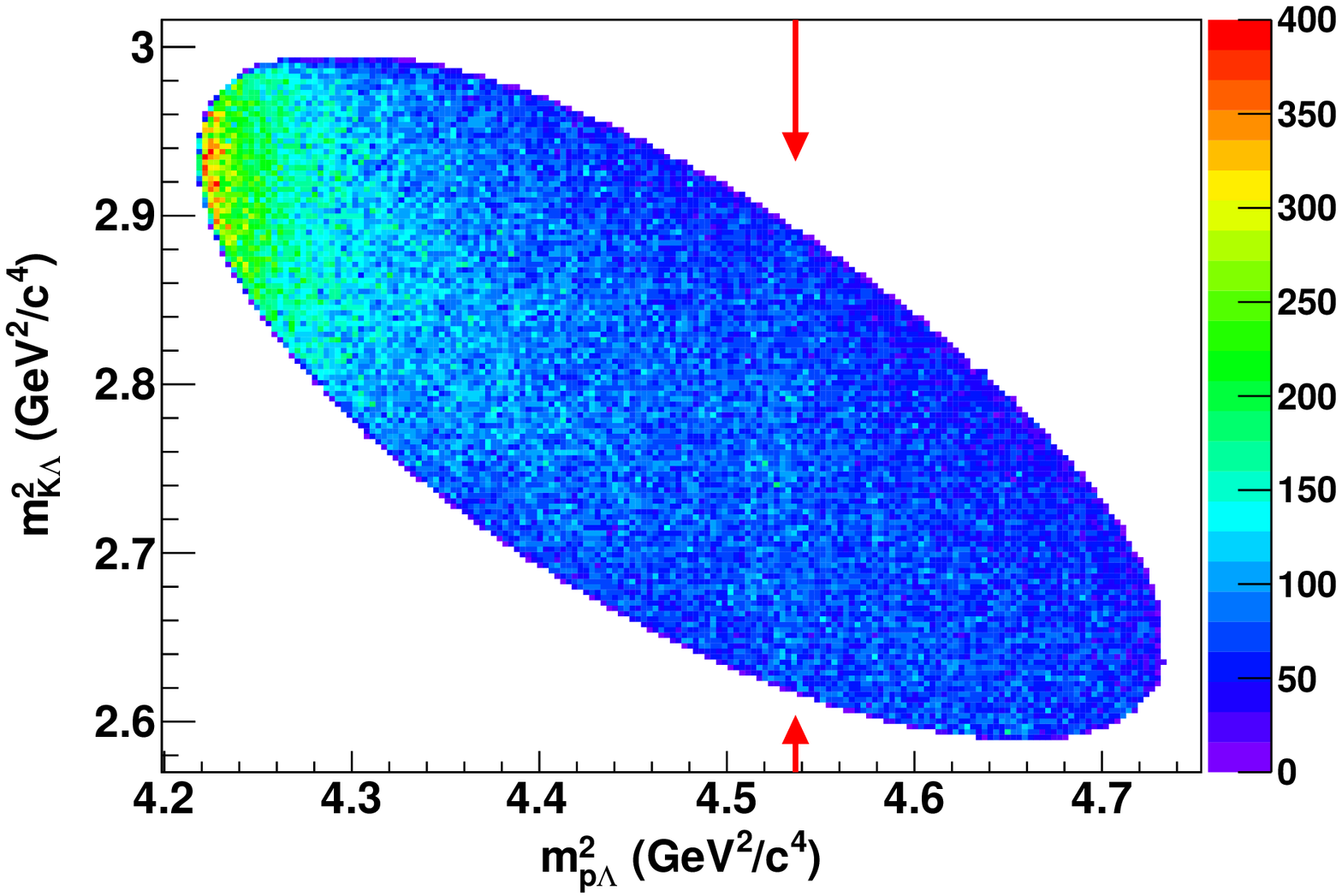}
\caption{\label{fig:dalitzplot}Dalitz plot of the selected event sample at $\SI{2.7}{\giga\eVperc}$ beam momentum corrected for acceptance and
reconstruction efficiency. The red arrows indicate the region of the $N\Sigma$ thresholds.}
\end{figure}

The Dalitz plot of the event sample is shown in Fig.~\ref{fig:dalitzplot}. It has been corrected for acceptance and reconstruction
efficiency by MC phase space generated events. The full kinematic acceptance of the COSY-TOF detector is evident. At low $p\Lambda$ masses a strong 
enhancement from the final state interaction is clearly visible. The
$N^{*}$ resonances that would be visible as horizontal bands do not appear due to their width of about $\SI{100}{\mega\eVperc\squared}$. However, they can distort the 
final-state interaction because off interferences which has been shown by means of a Dalitz plot analysis in Ref.~\cite{AbdEl-Samad2010}. Indeed, a deviation of 
about $\SI{1}{\femto\meter}$ on extracted scattering length values has been found in the analysis of a previous COSY-TOF at a higher beam momentum of $\SI{2.95}{\giga\eVperc}$ \cite{Roeder2013}. In the data presented here this effect is very small as discussed in detail later.

\begin{figure}[b]
\includegraphics[width=0.48\textwidth]{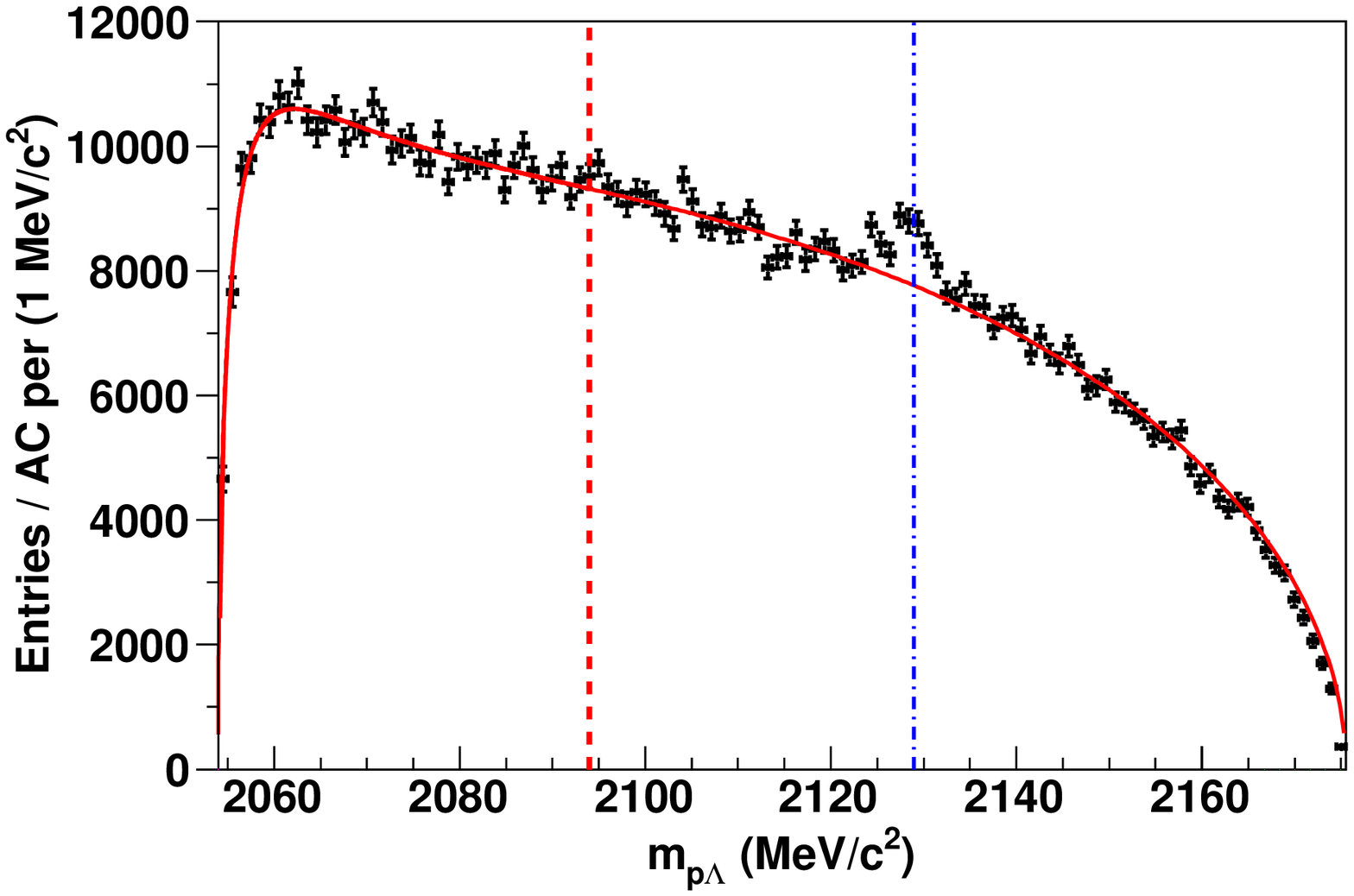}
\caption{\label{fig:projection}(color online). $p\Lambda$ invariant mass spectrum for the selected event sample at $\SI{2.7}{\giga\eVperc}$ beam momentum corrected for acceptance and
reconstruction efficiency (AC). The upper limit of the fit range is marked by the vertical dashed (red) line. The fit applied later (see Fig.~\ref{fig:mpldivspectrum}) and its continuation over the whole spectrum is shown by the solid (red) line. The vertical dash-dotted (blue) line indicates the lower $N\Sigma$ threshold ($n\Sigma^{+}$). }
\end{figure}

No pronounced enhancement at the $N\Sigma$ thresholds from the $N\Sigma-p\Lambda$ coupled
channel effect can be observed in Fig.~\ref{fig:dalitzplot}. However, the enhancement is clearly visible in the $p\Lambda$ invariant mass spectrum which is shown in Fig.~\ref{fig:projection} but it is weaker than in measurements at higher beam momenta \cite{Roeder2013,AbdEl-Samad2010} due to the available phase space at higher beam momentum. The enhancement does not distort the results obtained from the fit of the final state interaction as it was shown in a previous analysis of COSY-TOF data at higher beam momentum \cite{Jowzaee2016}. Furthermore, the fit of the final state interaction (see details below) extrapolated to the full
$m_{p\Lambda}$ range describes the spectrum well as shown by the solid line in Fig.~\ref{fig:projection}. 

\begin{figure}[tb]
\includegraphics[width=0.48\textwidth]{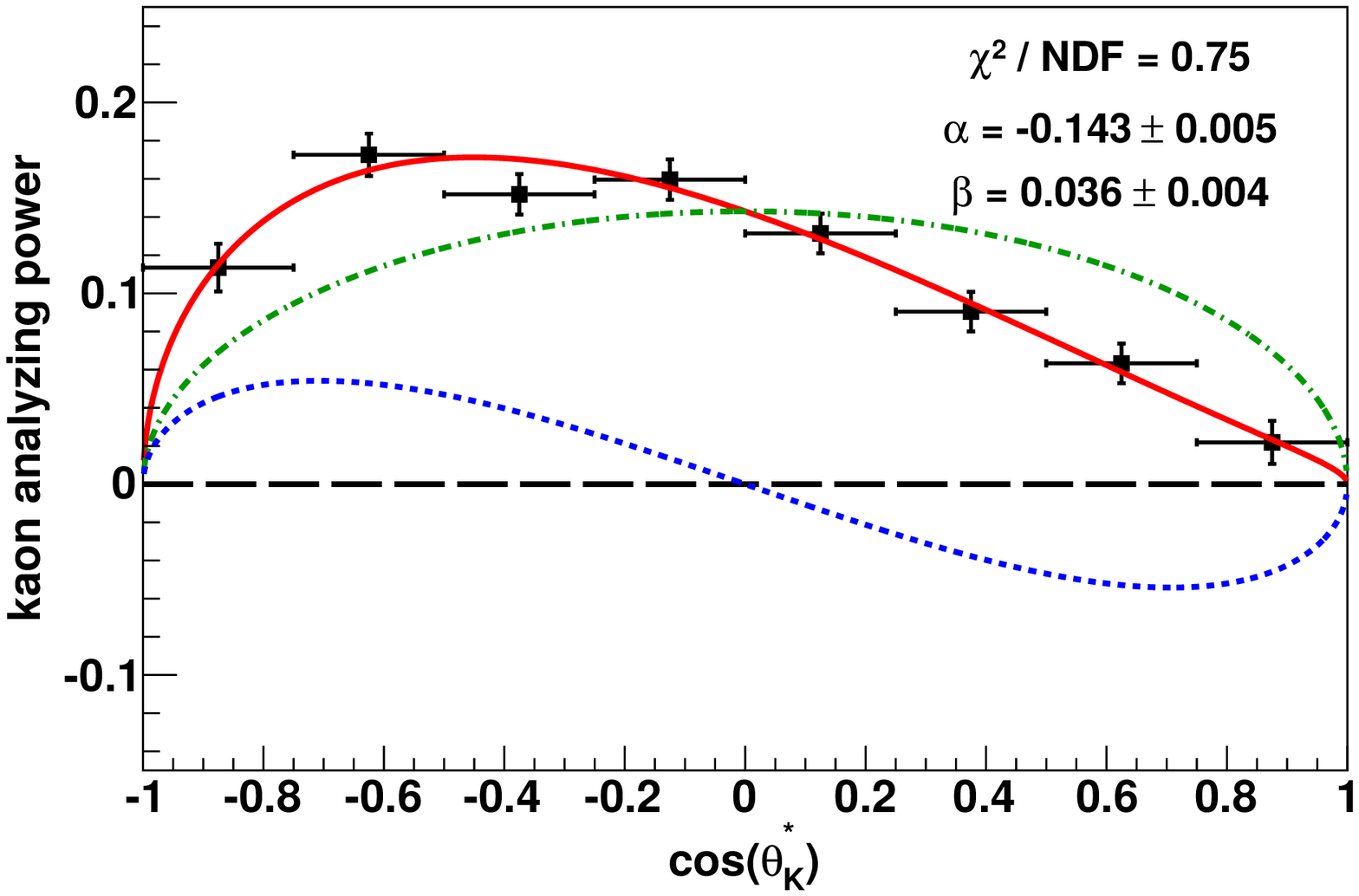}
\caption{\label{fig:kaonap}(color online). The beam analyzing power determined by the kaon asymmetry, $A_{N}^{K}$, as a function of $\cos(\theta_{K}^{*})$ for the full $m_{p\Lambda}$ range.
The solid (red) line shows the fit with $\alpha \cdot P_{1}^{1} + \beta \cdot P_{2}^{1}$. The individual contributions of the associated Legendre polynomials are
shown by the dash-dotted (green) and dotted (blue) lines, respectively.}
\end{figure}
In Fig.~\ref{fig:kaonap} the beam analyzing power determined by the kaon asymmetry, $A_{N}^{K}$, is shown as a function of $\cos(\theta_{K}^{*})$ for the full $m_{p\Lambda}$ range. The
fit with the associated Legendre polynomials $P_{1}^{1}$ and $P_{2}^{1}$ (solid line) reproduces the data within their statistical errors. The individual contributions of $P_{1}^{1}$ and $P_{2}^{1}$ are shown by the dash-dotted and dotted line,
respectively. Including higher order contributions does not improve the fit. These contributions are compatible with zero. 

\begin{figure}[tb]
\includegraphics[width=0.48\textwidth]{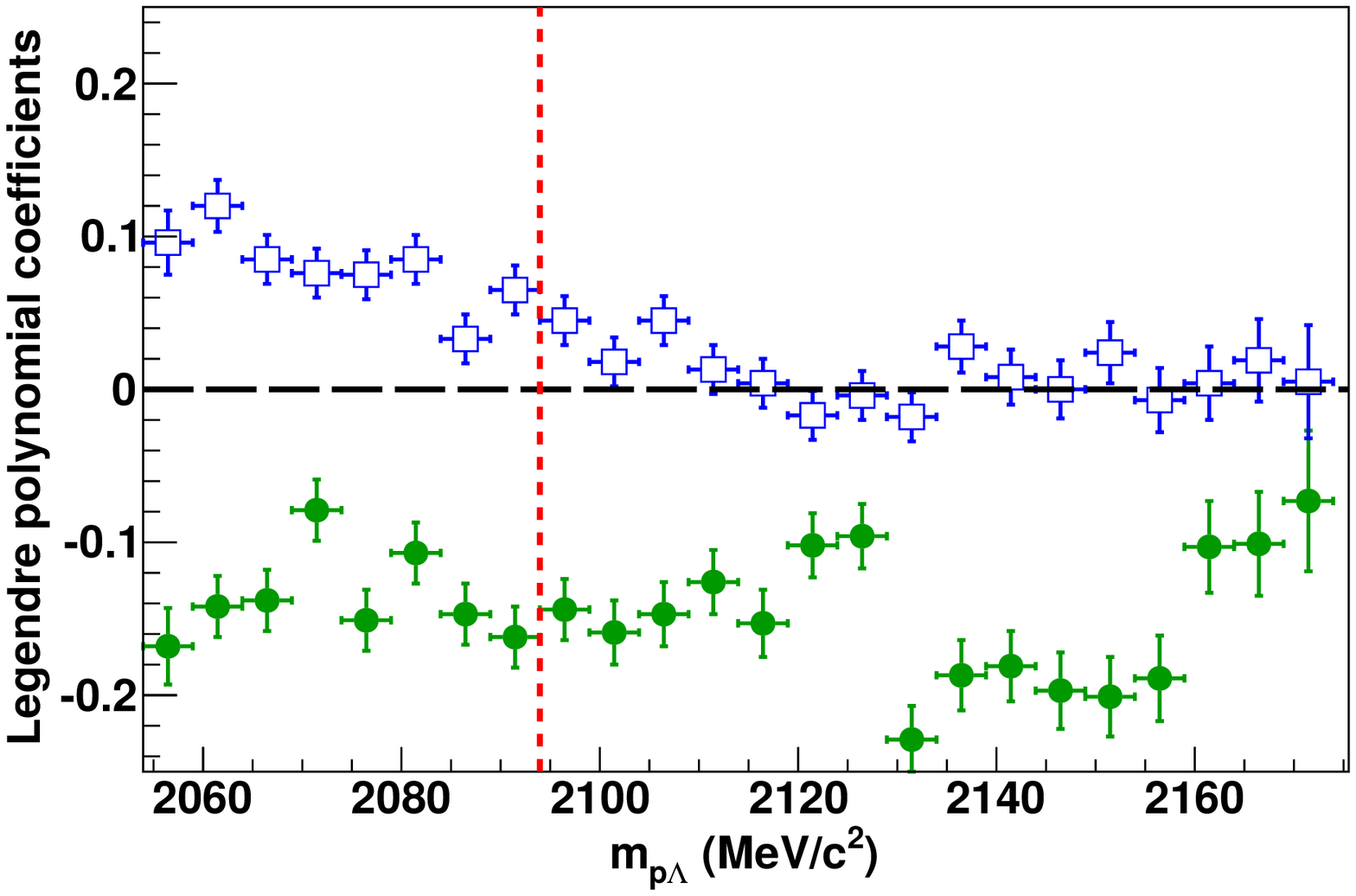}
 \caption{\label{fig:legendrecoeff}(color online). The coefficients $\alpha(m_{p\Lambda})$ (filled circles, green) and $\beta(m_{p\Lambda})$ (open squares, blue) from the fit of
  $A_{N}^{K}$ with associated Legendre polynomials as a function of $p\Lambda$ invariant mass. The dashed (red) vertical line indicates the 
  upper limit of the fitting range applied to the invariant mass spectrum.}
\end{figure}

In Fig.~\ref{fig:legendrecoeff} the coefficients $\alpha(m_{p\Lambda})$ (filled circles) and $\beta(m_{p\Lambda})$ (open squares) from the fit of $A_{N}^{K}$ are shown in
$\SI{5}{\mega\eVperc\squared}$ wide bins of $m_{p\Lambda}$. The right end of the spectrum corresponds to the kinematic limit at 
this beam momentum. There, both contributions have to be zero since the kaon has vanishing momentum and hence it has to be purely in S-wave without interference with higher order partial waves. It is interesting to note that $\alpha(m_{p\Lambda})$ changes significantly at the $N\Sigma$ thresholds at about $\SI{2130}{\mega\eVperc\squared}$ whereas $\beta(m_{p\Lambda})$ does not change.

The behavior of $\alpha(m_{p\Lambda})$ observed here is different from that in the measurement at higher beam momentum \cite{Roeder2013}, where 
$\alpha(m_{p\Lambda})$ was found to be compatible with zero at low invariant masses. In that case the extraction of the spin triplet scattering length was not possible with sufficient precision. A simple explanation for the vanishing $\alpha(m_{p\Lambda})$ value in that measurement \cite{Roeder2013} is a negligible production of the 
$p\Lambda$ system in the spin triplet state. Indeed, such a conclusion has been drawn in a paper of the HIRES collaboration \cite{Budzanowski2010a} from a combined
analysis of $p\Lambda$ elastic scattering cross sections and data from the $p\Lambda$ final state interaction in an inclusive $pp \rightarrow K^{+}{+}(\Lambda p)$ measurement at $\SI{2.735}{\giga\eVperc}$ beam momentum. However, this explanation is definitely excluded by the result shown in Fig.~\ref{fig:legendrecoeff}.

\begin{figure}[tb]
\includegraphics[width=0.48\textwidth]{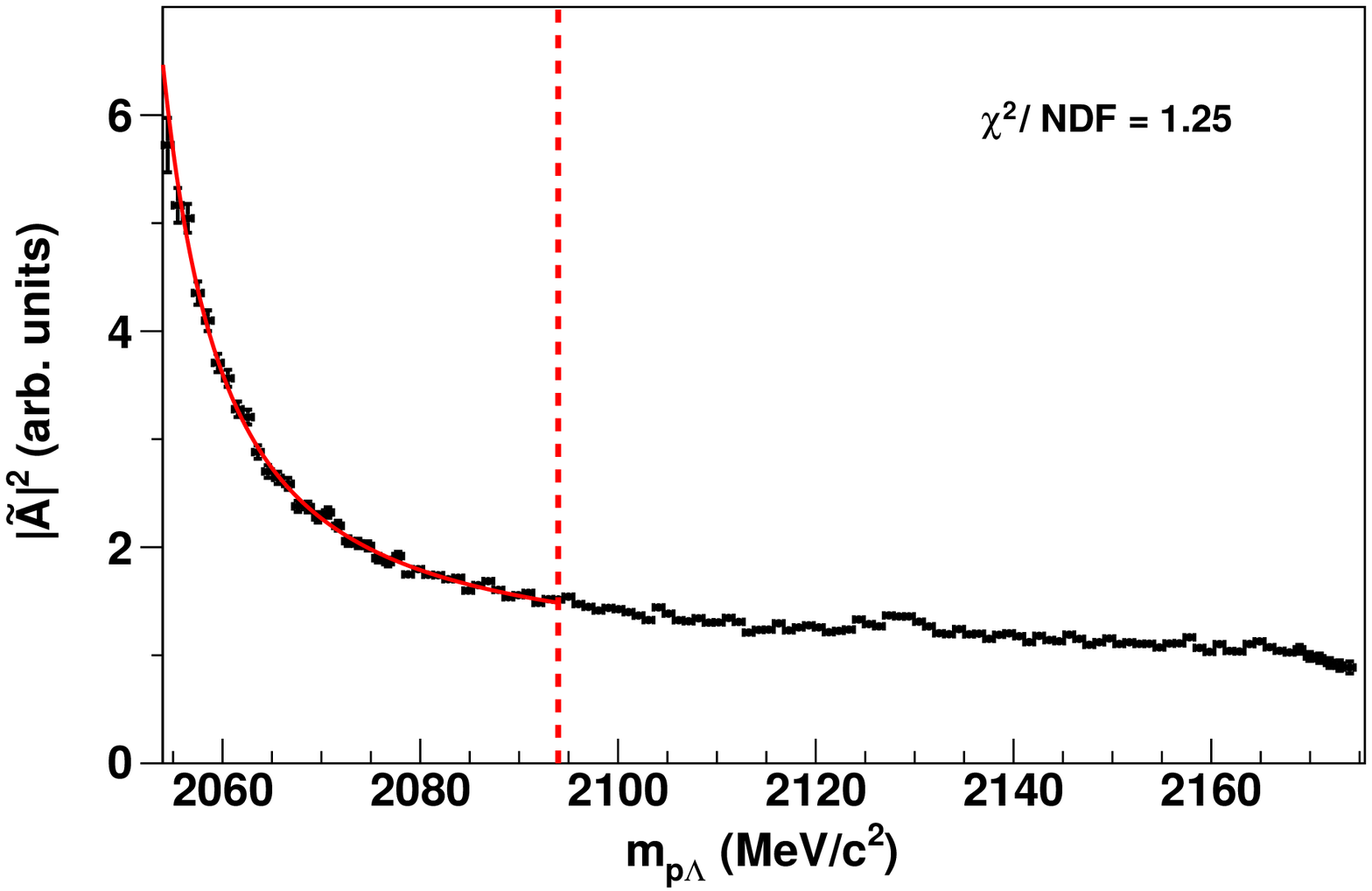}
\caption{\label{fig:mpldivspectrum}(color online). The amplitude squared $| \tilde{A}(m_{p\Lambda}) |^2$ corresponding to the $p\Lambda$ invariant mass
distribution divided by the volume of available phase space. The solid (red) line shows a fit to the data by Eq.~(\ref{eq:fsiamplitudeparametrization}). The fit values are $C_0 = -0.139^{+0.031}_{-0.033}$, $C_1 = \SI[parse-numbers=false]{(0.121 ^{+0.008}_{-0.007})}{\giga\eVperc\squared}$ and $C_2 = \SI[parse-numbers=false]{(4.158 ^{+0.003}_{-0.004})}{\giga\eVperc\squared}$. The vertical (red) dashed line marks the upper limit of the fitting range. }
\end{figure}

In a first step of the analysis, the unpolarized invariant mass distribution divided by the phase space, $| \tilde{A}(m_{p\Lambda}) |^2$, is fit using the parametrization (\ref{eq:fsiamplitudeparametrization}). From the fit the so-called effective $p\Lambda$ scattering length $a_{\mathrm{eff}}$ is calculated by Eq.~(\ref{eq:scatteringlengthformula}). The value is referred to be ``effective'' since the relative weights of the spin singlet and spin triplet final states are unknown for the unpolarized invariant mass distribution. It is not the spin averaged value determined in fits of $p\Lambda$ elastic scattering data. Nevertheless, the effective
value is determined for a comparison with previous analyses in \cite{Roeder2013,Gasparyan2004} and for the study of the influence of $N^{*}$ resonances on the scattering length value. Furthermore, systematic effects are studied for the effective scattering length due to the higher statistical precision in this case.
It is assumed that the systematic effects are the same for the obtained spin triplet value. 

Fig.~\ref{fig:mpldivspectrum} shows $| \tilde{A}(m_{p\Lambda}) |^2$ and its fit (solid line). The fit parameters and its asymmetric errors are given in the figure caption. The errors are calculated with the MINOS routine of the MINUIT2 library of the ROOT data analysis framework \cite{Brun1997a}. The dashed vertical line indicates the upper limit of 
the fit $m_{p} + m_{\Lambda} + \SI{40}{\mega\eV}$.
To calculate the scattering length and its error from the highly correlated parameters $C_i$ and their asymmetric errors, a bootstrapping method is used
with 5,000 simulations \cite{Vetterling2007,Efron1993}. From the scattering length distribution of the simulations, the value 
$a_{\mathrm{eff}} = \SI[parse-numbers=false]{(-1.38 ^{+0.04}_{-0.05} {}_{stat.} \pm 0.22_{syst.} \pm 0.3_{theo.})}{\femto\meter}$ is obtained. 
This value
is in agreement with the previous analysis in \cite{Roeder2013} as well as the analysis of \cite{Gasparyan2004} in which the theoretical uncertainty
is determined. The deviation to other methods treating FSI effects with the effective range approximation or Jost function are in the order of $\SI{0.6}{\femto\meter}$ (see Table I in \cite{Gasparyan2005}). 

Several systematic checks have been performed. The upper limit of the fit was varied between $\SI{35}{\mega\eV}$ and $\SI{60}{\mega\eV}$ above
threshold which gives similar results within the statistical error. Additionally, acceptance corrections were varied as well as binning issues were checked. Their contributions to the systematic error are $\SI{0.2}{\femto\meter}$ and $\SI{0.02}{\femto\meter}$, respectively.

Another systematic error results from the influence of
$N^{*}$ resonances, in particular the non S-wave resonances $N^{*}(1710)$ and $N^{*}(1720)$. To study the resonance effect, the Dalitz plot is sliced in four ranges of the $p\Lambda$ helicity angle $\cos\theta^{Rp\Lambda}_{pK}$.
Each slice occupies the full $m_{p\Lambda}$ range, and the invariant mass spectra for each slice would be practical identical without any $N^{*}$ contributions.
For each invariant mass 
spectrum a value for $a_{\mathrm{eff}}$ is determined. The root mean square deviation of these values is $\SI{0.1}{\femto\meter}$ and gives the error from the influence of $N^{*}$ resonances \cite{Hauenstein2014}.

\begin{figure}[tb]
\includegraphics[width=0.48\textwidth]{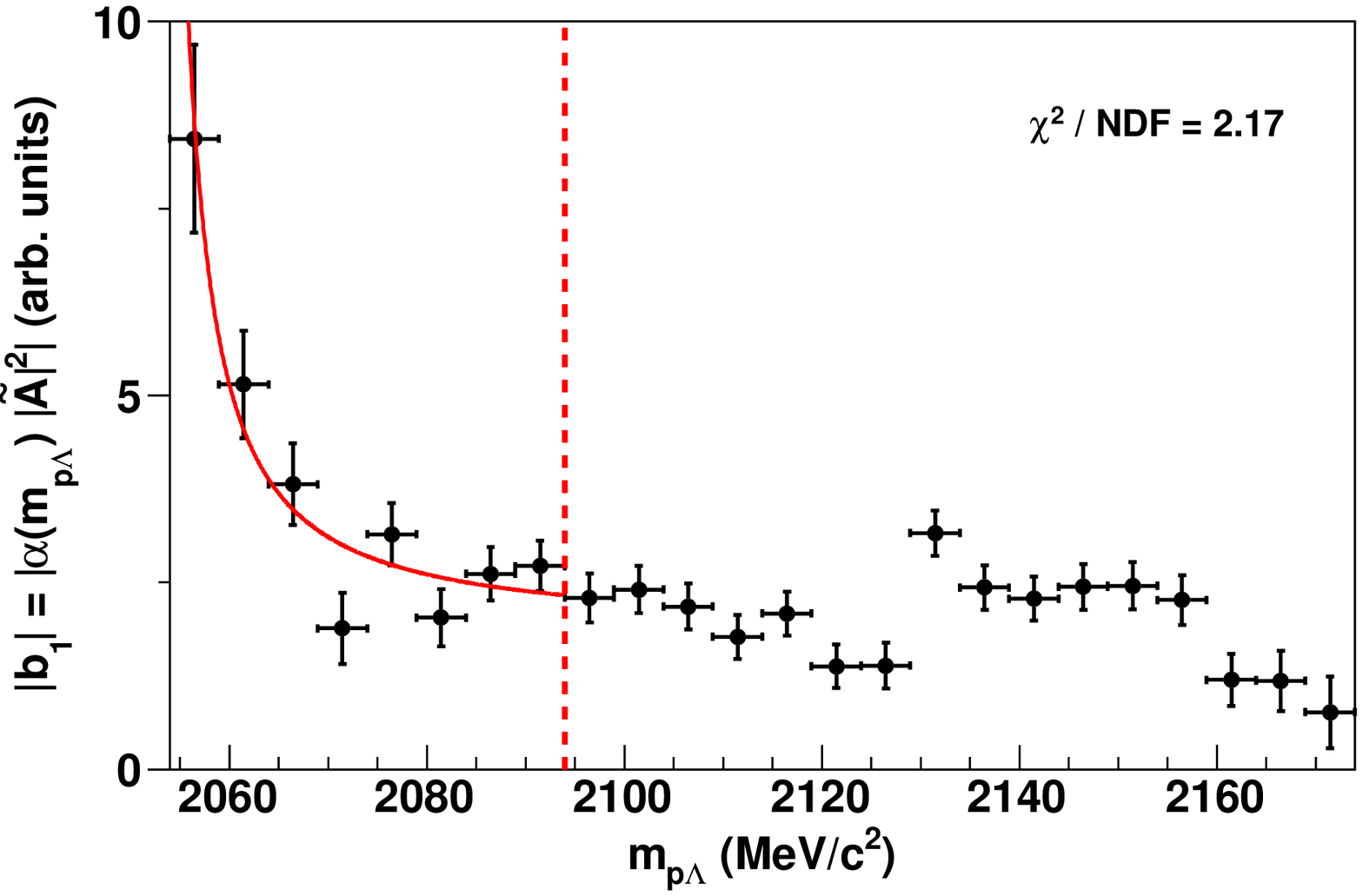}
\caption{\label{fig:mpldivspectrum2}(color online). The $|b_{1}(m_{p\Lambda})|$ distribution determined from $| \tilde{A}(m_{p\Lambda}) |^2$ and $|\alpha(m_{p\Lambda})|$. The solid (red) line shows the fit with the parametrization (\ref{eq:fsiamplitudeparametrization}) from which the 
spin triplet $p\Lambda$ scattering length is determined. The fit values are $C_0 = 0.599^{+0.171}_{-0.233}$, $C_1 = \SI[parse-numbers=false]{(0.046 ^{+0.035}_{-0.020})}{\giga\eVperc\squared}$ and $C_2 = \SI[parse-numbers=false]{(4.200^{+0.010}_{-0.017})}{\giga\eVperc\squared}$. The vertical (red) dashed line marks the upper limit of the fitting range.}
\end{figure}
In order to determine the spin triplet scattering length, $| \tilde{A}(m_{p\Lambda}) |^2$ is combined with $|\alpha(m_{p\Lambda})|$ to obtain
$|b_{1}(m_{p\Lambda})|$. In this case the spectrum, $| \tilde{A}(m_{p\Lambda}) |^2$, is rebinned to have sufficient statistics for the analyzing power in each $m_{p\Lambda}$ bin. The resulting distribution for $|b_{1}(m_{p\Lambda})|$ is
depicted in Fig.~\ref{fig:mpldivspectrum2}. The distribution is fitted according to Eq.~(\ref{eq:fsiamplitudeparametrization}), and the fit parameters and errors are given in the figure caption. They are calculated as before with the MINOS routine of the ROOT data analysis framework \cite{Brun1997a}. A bootstrapping method with 100,000 simulations is used to determine the value for the spin triplet scattering length and its statistical error \cite{Vetterling2007,Efron1993}. The result is $a_{t} = \SI[parse-numbers=false]{(-2.55 ^{+0.72}_{-1.39} {}_{stat.} \pm 0.6_{syst.} \pm 0.3_{theo.})}{\femto\meter}$. The larger systematic error stems from stronger binning effects compared to the determination of the effective scattering length.

\begin{table}[bt]
\caption{\label{tab:comparedata}Comparison of the spin triplet values from this paper with results from other experiments and theoretical predictions.}
\begin{ruledtabular}
\begin{tabular}{lllll}
{} & $a_t (\textrm{fm})$ & stat.(fm) & sys.(fm) & theo.(fm) \\\hline
this paper & $-2.55$ & ${}^{+0.72}_{-1.39}$ & $\pm 0.6 $ & $\pm 0.3 $  \\
$pp \rightarrow K^{+}{+}(\Lambda p)$ \cite{Budzanowski2010a}\footnote{Here, the inclusive data is fit together with the data from \cite{Alexander1968} 
in a combined, model-dependent procedure with the result from \cite{Tan1969} as a constraint.} & $-1.56$ & ${}^{+0.19}_{-0.22}$ & &  $\pm 0.4$  \\
$p\Lambda$ scattering \cite{Alexander1968} & $-1.6$ & ${}^{+1.1}_{-0.8}$ & & \\
$K^{-}d \rightarrow \pi^{-}p\Lambda$ \cite{Tan1969} & $-2.0$ & $\pm 0.5$ &  &\\\hline
$\chi$EFT NLO (500) \cite{Haidenbauer2013} & $-1.61$& & &\\
$\chi$EFT NLO (700) \cite{Haidenbauer2013} & $-1.48$& & &\\
J\"ulich 04 model \cite{Haidenbauer2005} & $-1.66$ & & & \\
Nijmegen NSC97f \cite{Rijken1999} & $-1.75$ & & &  \\

\end{tabular}
\end{ruledtabular}
\end{table}
Results for the spin triplet scattering length from experiments and theoretical predictions are shown in Table~\ref{tab:comparedata}.
The determined scattering length in this paper is compatible within $1\sigma$ of the statistical and theoretical uncertainty to the theoretical calculation
from next-to-leading order chiral effective field theory ($\chi$EFT)
\cite{Haidenbauer2013} and from meson-exchange based models, J\"ulich 04 \cite{Haidenbauer2005} and Nijmegen NSC97f \cite{Rijken1999}. All of these
calculations reproduce the measured hypertriton binding energy \cite{Nogga2014}.

The result from this paper is also compatible with the model-dependent analyses using the effective range approximation from \cite{Alexander1968,Tan1969}. As mentioned in the context of the analyzing power, the result from the HIRES collaboration \cite{Budzanowski2010a} stems from a combined fit of $p\Lambda$ elastic scattering data and an inclusively measured
 $p\Lambda$ invariant mass spectrum with the result of \cite{Tan1969} as a constraint. In this way a value for the spin
triplet scattering length has been obtained, although the authors of \cite{Budzanowski2010a} conclude a negligible production of spin triplet $p\Lambda$
states which is excluded by the results for the analyzing power shown in Fig.~\ref{fig:legendrecoeff}. 
Therefore, the result from this paper can not be compared with their result.

%
In summary, the $\vec{p}p \rightarrow pK^{+}\Lambda$ reaction has been measured with the COSY-TOF detector at a beam momentum of $\SI{2.7}{\giga\eVperc}$. Optimized
selection criteria give a data sample of $232,873$ kinematically fitted events.

For the determination of the spin triplet $p\Lambda$ scattering length from the final state interaction the beam analyzing power determined from the kaon asymmetry is evaluated in terms of
associated Legendre polynomials ($P_{1}^{1}$ and $P_{2}^{1}$) and the $p\Lambda$ invariant mass. The symmetric contribution $\alpha(m_{p\Lambda})$ to the analyzing power
is nearly constant and non zero for low invariant masses. Therefore, the extraction of the spin triplet scattering length is possible by the model independent method from Gasparyan et al. \cite{Gasparyan2004}. Furthermore, this result excludes the explanation of a dominant production of $p\Lambda$ spin singlet states in $pp \rightarrow pK^{+}\Lambda$ given
in an analysis of an inclusive measurement at a similar beam momentum \cite{Budzanowski2010a}.

The spin triplet $p\Lambda$ scattering length is obtained to be $a_{t} = \SI[parse-numbers=false]{(-2.55 ^{+0.72}_{-1.39} {}_{stat.} \pm 0.6_{syst.} \pm 0.3_{theo.})}{\femto\meter}$. This is the first direct determination of this parameter without relying on model assumptions or $p\Lambda$ elastic scattering data with mixed spin states \cite{Alexander1968,Budzanowski2010a}.
The systematic error from the influences of $N^{*}$ resonances has been studied
by analyzing different Dalitz plot slices. The error is of the order of $\SI{0.1}{\femto\meter}$ which is a factor ten smaller compared to a measurement at 
$p_{\mathrm{beam}} = \SI{2.95}{\giga\eVperc}$ \cite{Roeder2013}. This is in agreement with the expectation of a weaker influence of the non S-wave $N^{*}$ resonances, $N^{*}(1710)$ and $N^{*}(1720)$, on the production mechanism at lower beam momenta as it has been determined by a previous Dalitz plot analysis \cite{AbdEl-Samad2010}.
In addition, we want to point out that results from a combined partial wave analysis of several $pp \rightarrow pK\Lambda$ data sets will be published soon. This partial wave analysis addresses especially the influence of the different $N^{*}$ resonances in the production.

It is also necessary to investigate the effect of other polarization observables, e.g. the $\Lambda$ polarization, in order to put further constraints on the
production mechanism of associated strangeness as well as on the creation of $p\Lambda$ spin triplet states. These results are published in \cite{Hauenstein2016} but further theoretical considerations are necessary to set other constraints.

\begin{acknowledgements}

The authors want to thank the COSY crew for the excellent beam preparation, and
J. Uehlemann and N. Paul for the operation of the LH$_2$ target. 
Discussions with H. Machner and C. Wilkin are gratefully acknowledged. 
This work was supported by grants from Forschungszentrum J\"{u}lich (COSY-FFE),
by the European Union Seventh Framework program (FP7/2007-2013) under
grant agreement 283286.

\end{acknowledgements}

\bibliography{references}

\end{document}